\documentclass[aps,pre,reprint,superscriptaddress,floatfix]{revtex4-2}
\usepackage{amsmath,amssymb,bm,mathtools}
\usepackage{graphicx}
\usepackage{booktabs}
\usepackage{array}
\usepackage{siunitx}
\usepackage{hyperref}
\hypersetup{hidelinks}
\usepackage{xcolor}
\usepackage{enumitem}
\graphicspath{{figures/}}
\newcommand{\kb}{k_{\mathrm B}}
\newcommand{\ep}{\dot S_{\mathrm{tot}}}
\newcommand{\muu}{\Delta\mu}
\newcommand{\Th}{T_{\mathrm h}}
\newcommand{\Tc}{T_{\mathrm c}}
\newcommand{\fs}{f_{\mathrm{stall}}}
\newcommand{\Gc}{\Gamma}

\begin{document}

\title{Exact chemo--thermal Metropolis Brownian engine:\\
chemical leverage, temperature-neutral stall, power optimization, and multicyclic dissipation}

\author{Mesfin Asfaw Taye}
\email{tayem@wlac.edu}
\affiliation{Science Division, West Los Angeles College, 9000 Overland Avenue, Culver City, California 90230, USA}

\date{\today}

\begin{abstract}
We develop an exactly solvable chemo--thermal extension of the three-state Metropolis Brownian heat engine.  The particle moves through the periodic energy sequence $0\to E\to 2E\to0$, performs mechanical work against a load $f$ on every forward step, interacts with two cold links and one hot link, and consumes one fuel molecule of free-energy drop $\muu$ on the hot transition.  Local detailed balance gives an exact cycle affinity
\begin{equation*}
\mathcal A=E\left(\Tc^{-1}-\Th^{-1}\right)+\muu/\Th-f\left(2/\Tc+1/\Th\right),
\end{equation*}
and the full stationary probabilities and current are obtained without linear-response, weak-driving, or high-barrier approximations.  Several results follow.  First, the exact stall force is
\begin{equation*}
\fs=\frac{E(\Th-\Tc)+\Tc\muu}{2\Th+\Tc}.
\end{equation*}
Second, there is a temperature-neutral chemical compensation point $\muu_*=3E/2$ at which $\fs=E/2$ for every $\Th>\Tc$ and the hot and cold heats both vanish at reversible stall.  Third, in both Metropolis branches the stationary current is a strictly increasing function of $\muu$ at fixed mechanical parameters, but approaches a finite kinetic ceiling because chemical driving can remove only one of the three kinetic bottlenecks.  Fourth, an exact operating-mode map separates a genuine hybrid heat engine, a chemically driven motor dissipating into both baths, and a chemical motor that simultaneously pumps heat from cold to hot while delivering mechanical work.  Fifth, the maximum-power load obeys an exact scalar stationarity equation; close to equilibrium it reduces to $f_{\rm mp}=\fs/2$, while the corresponding exergy efficiency becomes $3/(6-\eta_C)$.  Sixth, for a general unicyclic network, chemical placement enters stall through the local gating temperature, implying a universal chemical-leverage factor proportional to $1/T_{\rm ch}$.  Finally, a minimal two-cycle extension demonstrates analytically that mechanical stall need not be thermodynamically reversible: a futile fuel cycle sustains positive entropy production while the mechanical velocity is zero.  The results connect Brownian heat engines and molecular motors within one transparent stochastic-thermodynamic model.
\end{abstract}

\maketitle

\section{Introduction}

Brownian heat engines and molecular motors are two closely related realizations of energy transduction at mesoscopic scales.  In the first, directed motion and work are generated from a temperature difference; in the second, chemical free energy supplied by nonequilibrium solute concentrations drives motion and force generation.  Both are naturally described by stochastic thermodynamics, in which transition rates obey local detailed balance and currents are conjugate to thermodynamic affinities \cite{Schnakenberg1976,Julicher1997,Seifert2011,Seifert2012,SchmiedlSeifert2007,RaoEsposito2016,Mugnai2020}.  For heat engines, the same framework has yielded exactly solvable models of efficiency at maximum power \cite{SchmiedlSeifert2008} and universal trade-offs between power, efficiency, and constancy \cite{Pietzonka2018}.  Chemical coupling to Brownian ratchets has a long history, including Magnasco's molecular combustion motor \cite{Magnasco1994}; hence the novelty sought here is not the generic statement that chemical energy can drive a ratchet, but the exact chemo--thermal solution and the new identities that follow for the particular three-state Metropolis heat engine.

Discrete-state Brownian heat engines operating between two reservoirs have a long record as minimal, analytically tractable models of energy conversion \cite{Asfaw2002AdjustableBrownianHeat,Asfaw2004CurrentMaximumPower,Asfaw2005EnergeticsSimpleMicroscopic,Asfaw2007ExploringOperationTiny,Asfaw2008ModelingEfficientBrownian,Asfaw2014ThermodynamicFeatureBrownian}.  The purely thermal three-state version admits unusually transparent analytical thermodynamics.  Exact expressions for current, efficiency, entropy production, and related quantities have been developed for piecewise, linear, quadratic, and exponential temperature architectures \cite{Taye2015ExactAnalyticalThermodynamic,Taye2016FreeEnergyEntropy,Taye2017IrreversibleBrownianHeat,Taye2020EntropyProductionEntropy,Taye2022ExactTimeDependent2,Taye2023TimeDependentSolutions,Taye2024ExactTimeDependent,Taye2025CurzonAhlbornType}, including underdamped \cite{Taye2021EffectViscousFriction,Taye2025ThermodynamicIrreversibilityUnderdamped}, multi-particle and polymeric \cite{Taye2015EffectTemperatureDependence,Taye2021BrownianMotorsArranged}, and hybrid active--passive \cite{Taye2025CompetingActivePassive,Taye2026ExactThermodynamicAnalysis} extensions; a systematic overview is given in Ref.~\cite{Taye2026BrownianMotorsBrownian}.  In particular, recent work showed that changing the spatial temperature architecture can strongly alter velocity, entropy production, and efficiency even when the endpoint temperatures are fixed \cite{Taye2024ExactTimeDependent,Taye2025CurzonAhlbornType}.  A natural next question is what happens if the same mesoscopic engine is powered simultaneously by a thermal gradient and a chemical free-energy source.

The present work answers that question in an exactly solvable setting.  We couple one fuel-consuming event to the hot transition of the three-state Metropolis engine.  The network remains unicyclic, so its stationary state can be solved in closed form and its cycle affinity controls both the direction of motion and the entropy produced per net cycle.  This provides a clean bridge between a Brownian heat engine and an isothermal chemical motor.

Beyond the basic stall shift, the combined model contains several results that are not visible in the passive engine.  We derive: (i) the exact stationary probabilities, not only the cycle current; (ii) an exact proof that chemical driving monotonically accelerates the motor but saturates at a kinetic ceiling; (iii) a temperature-neutral compensation point $\muu_*=3E/2$ where the reversible stall load and work become purely chemical; (iv) a complete mode diagram; (v) exact maximum-power and barrier-optimization equations; (vi) a general theorem showing how the temperature of the chemically gated link controls chemical mechanical leverage; and (vii) a minimal multicyclic extension in which mechanical stall coexists with finite fuel consumption and finite entropy production, thereby removing a familiar limitation of unicyclic motor models \cite{Mugnai2020}.

Throughout we set $\kb=1$, the lattice spacing equal to unity, and use a common attempt frequency $\Gc$.  All energies, temperatures, and chemical potentials therefore carry the same units.

\section{Three-state chemo--thermal Metropolis model}

\subsection{States, load, reservoirs, and chemical event}

\begin{figure}[t]
\includegraphics[width=\columnwidth]{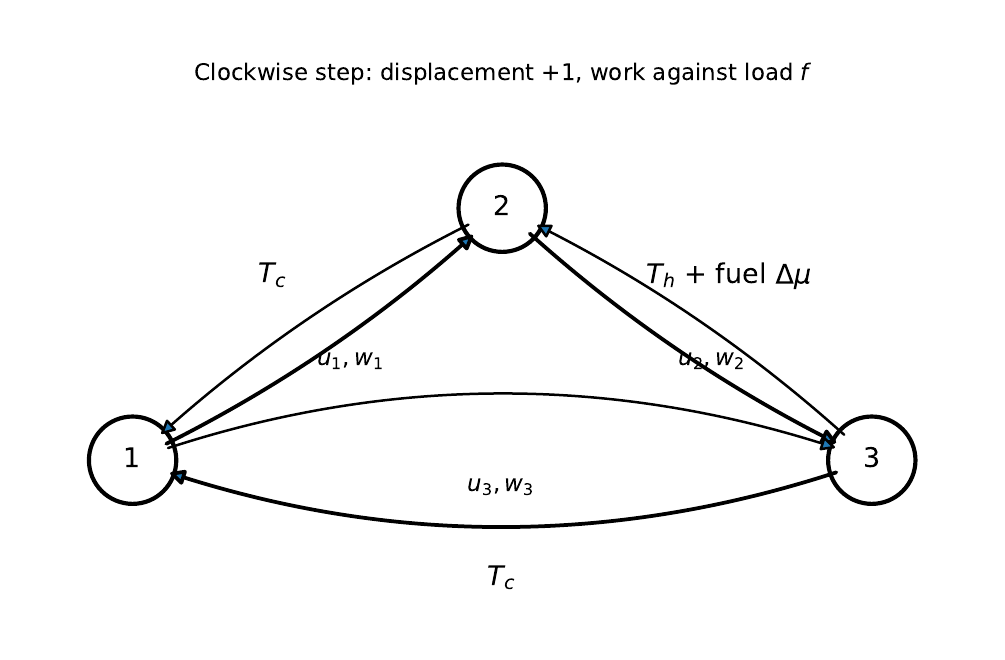}
\caption{Three-state chemo--thermal Metropolis engine.  The cold bath acts on links $1\leftrightarrow2$ and $3\leftrightarrow1$; the hot bath acts on $2\leftrightarrow3$.  A forward hot transition consumes one fuel molecule of free-energy drop $\muu$.  Each clockwise jump advances the particle by one lattice spacing against load $f$.}
\label{fig:model}
\end{figure}

Consider the periodic three-state sequence
\begin{equation}
1\longrightarrow2\longrightarrow3\longrightarrow1,
\end{equation}
with internal energies
\begin{equation}
\varepsilon_1=0,\qquad \varepsilon_2=E,\qquad \varepsilon_3=2E,
\label{energies}
\end{equation}
where $E>0$.  Every clockwise transition advances the particle by one lattice spacing against an opposing load $f\ge0$.  Thus a completed clockwise cycle advances the particle by three spacings and delivers the mechanical work
\begin{equation}
W=3f.
\label{workcycle}
\end{equation}
The links $1\leftrightarrow2$ and $3\leftrightarrow1$ are coupled to the cold reservoir $\Tc$, while $2\leftrightarrow3$ is coupled to the hot reservoir $\Th>\Tc$; see Fig.~\ref{fig:model}.

A fuel reaction is coupled stoichiometrically to the hot link.  During a clockwise $2\to3$ transition one fuel molecule is consumed and supplies the free energy $\muu>0$; the reverse $3\to2$ transition synthesizes the fuel.  The thermal energy increments that must be supplied to the system by the local baths during a clockwise jump are therefore
\begin{align}
\delta_1&=E+f,\label{delta1}\\
\delta_2&=E+f-\muu,\label{delta2}\\
\delta_3&=-2E+f.\label{delta3}
\end{align}
The sign convention is important: $\delta_i>0$ means that the system absorbs heat from the bath on the forward transition; $\delta_i<0$ means that it releases heat to that bath.

\subsection{Thermodynamically consistent Metropolis rates}

For a transition with thermal increment $\delta_i$ and local temperature $T_i$, we choose the Metropolis pair
\begin{align}
u_i&=\Gc\min\left\{1,e^{-\delta_i/T_i}\right\},\label{metu}\\
w_i&=\Gc\min\left\{1,e^{+\delta_i/T_i}\right\}.\label{metw}
\end{align}
These rates satisfy the local-detailed-balance identity
\begin{equation}
\ln\frac{u_i}{w_i}=-\frac{\delta_i}{T_i}
\label{ldb}
\end{equation}
for either sign of $\delta_i$.  The three ratios are therefore
\begin{align}
\frac{u_1}{w_1}&=\exp\left[-\frac{E+f}{\Tc}\right],\label{ratio1}\\
\frac{u_2}{w_2}&=\exp\left[-\frac{E+f-\muu}{\Th}\right],\label{ratio2}\\
\frac{u_3}{w_3}&=\exp\left[+\frac{2E-f}{\Tc}\right].\label{ratio3}
\end{align}
No assumption about which Metropolis branch is saturated is required for Eqs.~(\ref{ratio1})--(\ref{ratio3}).

\section{Master equation and exact stationary solution}

Let $p_i(t)$ denote the probability of state $i$.  With clockwise rates $u_i$ and counterclockwise rates $w_i$, the master equation is
\begin{align}
\dot p_1&=-(u_1+w_3)p_1+w_1p_2+u_3p_3,\label{me1}\\
\dot p_2&=u_1p_1-(w_1+u_2)p_2+w_2p_3,\label{me2}\\
\dot p_3&=w_3p_1+u_2p_2-(u_3+w_2)p_3,\label{me3}
\end{align}
with $p_1+p_2+p_3=1$.

At stationarity, the exact probabilities are
\begin{align}
p_1^{\rm s}&=\frac{u_2u_3+u_3w_1+w_1w_2}{D},\label{p1}\\
p_2^{\rm s}&=\frac{u_1u_3+u_1w_2+w_2w_3}{D},\label{p2}\\
p_3^{\rm s}&=\frac{u_1u_2+u_2w_3+w_1w_3}{D},\label{p3}
\end{align}
where
\begin{align}
D={}&u_1u_2+u_1u_3+u_1w_2+u_2u_3+u_2w_3+u_3w_1\notag\\
&+w_1w_2+w_1w_3+w_2w_3>0.
\label{Ddef}
\end{align}
A derivation by cofactors of the stationary generator is given in Appendix~\ref{app:stationary}.

The stationary edge currents coincide,
\begin{equation}
J=u_1p_1^{\rm s}-w_1p_2^{\rm s}
=u_2p_2^{\rm s}-w_2p_3^{\rm s}
=u_3p_3^{\rm s}-w_3p_1^{\rm s},
\end{equation}
and direct substitution yields
\begin{equation}
\boxed{
J=\frac{u_1u_2u_3-w_1w_2w_3}{D}.}
\label{Jexact}
\end{equation}
The mean lattice velocity is
\begin{equation}
V=3J.
\label{velocity}
\end{equation}

\section{Cycle affinity, current direction, and exact stall}

The dimensionless cycle affinity is
\begin{equation}
\mathcal A=\ln\frac{u_1u_2u_3}{w_1w_2w_3}.
\label{affdef}
\end{equation}
Using local detailed balance gives
\begin{align}
\mathcal A
&=-\frac{E+f}{\Tc}-\frac{E+f-\muu}{\Th}+\frac{2E-f}{\Tc}\label{affexpand}\\
&=E\left(\frac1{\Tc}-\frac1{\Th}\right)
+\frac{\muu}{\Th}
-f\left(\frac2{\Tc}+\frac1{\Th}\right).
\label{affinity}
\end{align}
It is convenient to define
\begin{equation}
\mathcal A_0=E(\Tc^{-1}-\Th^{-1})+\muu/\Th,
\qquad
C_f=2/\Tc+1/\Th,
\label{A0Cf}
\end{equation}
so that
\begin{equation}
\mathcal A=\mathcal A_0-C_ff=C_f(\fs-f).
\label{Astallform}
\end{equation}
Since
\begin{equation}
u_1u_2u_3-w_1w_2w_3=w_1w_2w_3(e^{\mathcal A}-1),
\end{equation}
Eq.~(\ref{Jexact}) immediately gives
\begin{equation}
\operatorname{sgn}J=\operatorname{sgn}\mathcal A.
\label{signJ}
\end{equation}
Consequently, the exact stall condition is $\mathcal A=0$:
\begin{equation}
\boxed{
\fs=\frac{E(\Th-\Tc)+\Tc\muu}{2\Th+\Tc}.}
\label{fstall}
\end{equation}
No diagonalization, weak-driving approximation, or expansion in $\Th-\Tc$ or $\muu$ is involved.

\subsection{Why unicyclic stall is reversible}

At stall, the product of the forward/backward ratios is unity,
\begin{equation}
\frac{u_1}{w_1}\frac{u_2}{w_2}\frac{u_3}{w_3}=1.
\label{productone}
\end{equation}
One may therefore construct stationary probabilities satisfying the three pairwise detailed-balance relations
\begin{equation}
u_1p_1=w_1p_2,\qquad
u_2p_2=w_2p_3,\qquad
u_3p_3=w_3p_1.
\label{dbstall}
\end{equation}
The product consistency condition is precisely Eq.~(\ref{productone}).  Hence each edge current vanishes separately.  A single-cycle network therefore cannot sustain hidden dissipation at mechanical stall.  This limitation motivates the multicyclic extension in Sec.~\ref{sec:multicycle}.

\section{Energy balance, entropy production, and generalized Carnot relation}

For one completed clockwise cycle the hot bath supplies
\begin{equation}
q_h=E+f-\muu,
\label{qh}
\end{equation}
while the net heat delivered to the cold reservoir is
\begin{equation}
q_c=E-2f.
\label{qc}
\end{equation}
The first law is the identity
\begin{equation}
\boxed{\muu+q_h=W+q_c.}
\label{firstlaw}
\end{equation}
The environmental entropy change per clockwise cycle is
\begin{align}
\Delta S_{\rm cyc}
&=\frac{q_c}{\Tc}-\frac{q_h}{\Th}\label{Scyc1}\\
&=\frac{E-2f}{\Tc}-\frac{E+f-\muu}{\Th}=\mathcal A.
\label{Scyc2}
\end{align}
At a steady state the system Shannon entropy is constant, and the total entropy-production rate is
\begin{equation}
\boxed{\ep=J\mathcal A\ge0.}
\label{epJA}
\end{equation}
A direct edgewise proof is
\begin{align}
\ep&=\sum_{i=1}^{3}J\ln\frac{u_ip_i^{\rm s}}{w_ip_{i+1}^{\rm s}}\notag\\
&=J\sum_i\ln\frac{u_i}{w_i}
+J\ln\frac{p_1p_2p_3}{p_2p_3p_1}=J\mathcal A.
\end{align}
Thus the second law follows immediately from Eq.~(\ref{signJ}).

Eliminating $q_c$ between the first and second laws gives
\begin{align}
\Tc\Delta S_{\rm cyc}
&=q_c-\frac{\Tc}{\Th}q_h\notag\\
&=\muu+q_h-W-\frac{\Tc}{\Th}q_h.
\end{align}
Therefore, with $\eta_C=1-\Tc/\Th$,
\begin{equation}
\boxed{
\muu+\eta_Cq_h-W=\Tc\Delta S_{\rm cyc}.}
\label{exergyidentity}
\end{equation}
For a forward-running cycle, $\mathcal A\ge0$, so
\begin{equation}
W\le\muu+\eta_Cq_h.
\label{generalbound}
\end{equation}
At reversible stall,
\begin{equation}
\boxed{W_{\rm rev}=\muu+\eta_Cq_h.}
\label{generalcarnot}
\end{equation}
The chemical term is work-quality free energy and therefore enters without a Carnot factor; the Carnot factor multiplies only the heat drawn from the hot reservoir.

A natural exergy efficiency for any forward work-producing regime with a positive net exergy input is
\begin{equation}
\boxed{
\eta_x=\frac{W}{\muu+\eta_Cq_h}
=\frac{3f}{\muu+\eta_C(E+f-\muu)}\le1.}
\label{etax}
\end{equation}
A naive ``thermal efficiency'' $W/q_h$ can exceed $\eta_C$ when chemical free energy contributes to the output; this does not violate the second law because Eq.~(\ref{generalbound}) rather than $W\le\eta_Cq_h$ is the correct bound.

\subsection{Quasistatic limits: thermal, chemical, and hybrid driving}
\label{sec:quasistatic}

The reversible limit of the present finite-state engine is the approach to stall from the forward-running side.  This point deserves an explicit derivation because the thermal and chemical resources have different reversible efficiencies.  Define the positive distance from stall
\begin{equation}
\epsilon=\fs-f>0,
\qquad \epsilon\rightarrow0^+ .
\label{epsstall}
\end{equation}
Using Eq.~(\ref{affinity}) together with Eq.~(\ref{fstall}), the affinity can be written exactly as
\begin{equation}
\boxed{
\mathcal A=C_f\epsilon,
\qquad
C_f=\frac{2}{\Tc}+\frac{1}{\Th}>0.}
\label{Aeps}
\end{equation}
Thus the quasistatic limit is equivalently
\begin{equation}
f\rightarrow\fs^{-}
\Longleftrightarrow
\mathcal A\rightarrow0^+ .
\label{qsequiv}
\end{equation}
The exact current in Eq.~(\ref{Jexact}) may be written as
\begin{equation}
J=\frac{w_1w_2w_3\left(e^{\mathcal A}-1\right)}{D}.
\label{Jaffnearstall}
\end{equation}
All rates and $D$ remain finite and positive at stall.  Hence, expanding $e^{\mathcal A}-1=\mathcal A+O(\mathcal A^2)$,
\begin{equation}
J=\mathcal L_{\rm st}\mathcal A+O(\mathcal A^2),
\qquad
\mathcal L_{\rm st}=\left.\frac{w_1w_2w_3}{D}\right|_{f=\fs}>0.
\label{Jlinearstall}
\end{equation}
Combining Eqs.~(\ref{Aeps}) and (\ref{Jlinearstall}) gives
\begin{equation}
J=\mathcal L_{\rm st}C_f\epsilon+O(\epsilon^2).
\label{Jeps}
\end{equation}
Consequently the mean velocity, mechanical power, and entropy-production rate behave as
\begin{align}
V&=3J=O(\epsilon),\label{Vqs}\\
\dot W&=3fJ=O(\epsilon),\label{Pqs}\\
\ep&=J\mathcal A
=\mathcal L_{\rm st}C_f^2\epsilon^2+O(\epsilon^3).
\label{EPqs}
\end{align}
Therefore reversible operation is reached with vanishing current and vanishing power, while the entropy-production rate vanishes one order faster.  This is the precise quasistatic limit of the unicyclic chemo--thermal engine.

\subsubsection{Pure thermal quasistatic limit}

Set $\muu=0$ while retaining $\Th>\Tc$.  Equation~(\ref{fstall}) becomes
\begin{equation}
\fs^{\rm th}
=\frac{E(\Th-\Tc)}{2\Th+\Tc}.
\label{fstthermalqs}
\end{equation}
The reversible mechanical work per cycle is therefore
\begin{equation}
W_{\rm rev}^{\rm th}=3\fs^{\rm th}
=\frac{3E(\Th-\Tc)}{2\Th+\Tc}.
\label{Wthermalqs}
\end{equation}
For $\muu=0$, Eq.~(\ref{qh}) gives $q_h=E+f$.  Evaluated at stall,
\begin{align}
q_{h,{\rm st}}^{\rm th}
&=E+\frac{E(\Th-\Tc)}{2\Th+\Tc}\notag\\
&=\frac{3E\Th}{2\Th+\Tc}.
\label{qhthermalqs}
\end{align}
Hence the ordinary heat-engine efficiency in the reversible limit is
\begin{align}
\eta_{\rm th}^{\rm qs}
&=\frac{W_{\rm rev}^{\rm th}}{q_{h,{\rm st}}^{\rm th}}\notag\\
&=\frac{3E(\Th-\Tc)/(2\Th+\Tc)}{3E\Th/(2\Th+\Tc)}\notag\\
&=\boxed{1-\frac{\Tc}{\Th}=\eta_C}.
\label{etathqs}
\end{align}
Thus the passive Brownian heat engine recovers the Carnot limit only quasistatically, where $J$, $V$, and $\dot W$ all vanish.

\subsubsection{Pure chemical quasistatic limit}

Now set the two reservoir temperatures equal,
\begin{equation}
\Th=\Tc=T.
\label{isothermalqs}
\end{equation}
The thermal part of the affinity cancels and Eq.~(\ref{affinity}) reduces to
\begin{equation}
\mathcal A_{\rm chem}=\frac{\muu-3f}{T}.
\label{Achemqs}
\end{equation}
Reversible stall therefore occurs at
\begin{equation}
\boxed{\fs^{\rm chem}=\frac{\muu}{3}.}
\label{fstchemqs}
\end{equation}
Since $W=3f$, the reversible work is
\begin{equation}
\boxed{W_{\rm rev}^{\rm chem}=\muu.}
\label{Wchemqs}
\end{equation}
The appropriate chemical conversion efficiency is
\begin{equation}
\eta_{\rm chem}=\frac{W}{\muu},
\end{equation}
so that
\begin{equation}
\boxed{\eta_{\rm chem}^{\rm qs}=1.}
\label{etachemqs}
\end{equation}
This unit reversible efficiency is not a Carnot statement.  In the isothermal limit $\eta_C=0$; the useful work is supplied entirely by chemical free energy, which is already a work-quality thermodynamic resource.

\subsubsection{Hybrid chemo--thermal quasistatic limit}

When both $\Th>\Tc$ and $\muu>0$, the exact stall load remains
\begin{equation}
\fs=\frac{E(\Th-\Tc)+\Tc\muu}{2\Th+\Tc}.
\label{fsthybridqs}
\end{equation}
At quasistatic stall, $\mathcal A\rightarrow0$ and therefore $\Delta S_{\rm cyc}\rightarrow0$.  Equation~(\ref{exergyidentity}) then becomes
\begin{equation}
\boxed{W_{\rm rev}=\muu+\eta_Cq_h.}
\label{Whybridqs}
\end{equation}
Thus the reversible work is the sum of the full chemical free energy and the Carnot-convertible fraction of the absorbed heat.  Correspondingly, the exergy efficiency defined in Eq.~(\ref{etax}) obeys
\begin{equation}
\boxed{
\eta_x^{\rm qs}
=\lim_{f\to\fs^-}
\frac{W}{\muu+\eta_Cq_h}=1.}
\label{etaxqs}
\end{equation}
Equations~(\ref{etathqs}), (\ref{etachemqs}), and (\ref{etaxqs}) display the three reversible limits in a common form: thermal work conversion approaches Carnot efficiency, chemical free-energy conversion approaches unit efficiency, and combined chemo--thermal conversion approaches unit exergy efficiency.

A particularly transparent member of the hybrid quasistatic family is the compensation point derived below.  At $\muu=3E/2$ one has $\fs=E/2$ and, at reversible stall, $q_h=q_c=0$.  Equation~(\ref{Whybridqs}) then reduces to $W_{\rm rev}=\muu$, despite $\Th\ne\Tc$.

\section{Chemical compensation point and exact stall sensitivities}
\label{sec:compensation}

\begin{figure}[t]
\includegraphics[width=\columnwidth]{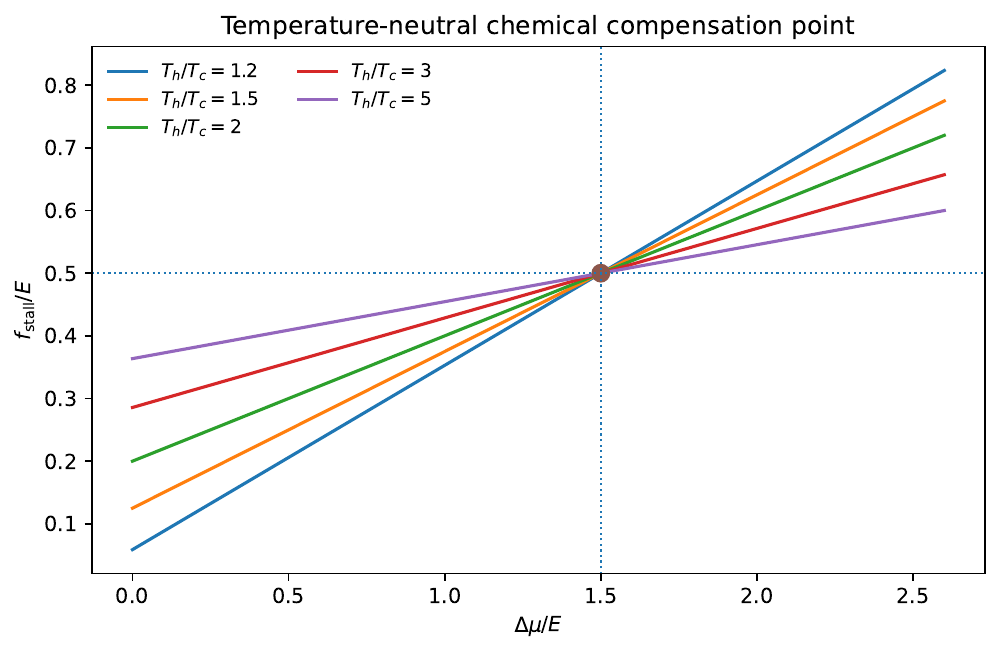}
\caption{Exact stall force versus chemical free energy for several hot/cold temperature ratios, with $E=\Tc=1$.  All curves cross at the temperature-neutral point $\muu/E=3/2$, $\fs/E=1/2$.}
\label{fig:compensation}
\end{figure}

Equation~(\ref{fstall}) contains a nontrivial crossover.  Differentiation gives
\begin{align}
\frac{\partial\fs}{\partial\muu}&=\frac{\Tc}{2\Th+\Tc}>0,\label{dfs_mu}\\
\frac{\partial\fs}{\partial E}&=\frac{\Th-\Tc}{2\Th+\Tc}>0,\label{dfs_E}\\
\frac{\partial\fs}{\partial\Th}&=
\frac{\Tc(3E-2\muu)}{(2\Th+\Tc)^2},\label{dfs_Th}\\
\frac{\partial\fs}{\partial\Tc}&=
-\frac{\Th(3E-2\muu)}{(2\Th+\Tc)^2}.
\label{dfs_Tc}
\end{align}
Both temperature derivatives vanish simultaneously at
\begin{equation}
\boxed{\muu_*=\frac{3E}{2}.}
\label{mustar}
\end{equation}
Substitution into Eq.~(\ref{fstall}) yields
\begin{equation}
\boxed{\fs(\muu_*)=\frac{E}{2},}
\label{fstar}
\end{equation}
independent of either reservoir temperature.  Figure~\ref{fig:compensation} shows the universal intersection of stall curves for several $\Th/\Tc$.

The physical meaning becomes clearer by evaluating the cycle heats at stall.  Using Eq.~(\ref{fstall}),
\begin{align}
q_h^{\rm st}&=E+\fs-\muu
=\frac{\Th(3E-2\muu)}{2\Th+\Tc},\label{qhstall}\\
q_c^{\rm st}&=E-2\fs
=\frac{\Tc(3E-2\muu)}{2\Th+\Tc}.
\label{qcstall}
\end{align}
Hence
\begin{equation}
\frac{q_c^{\rm st}}{\Tc}=\frac{q_h^{\rm st}}{\Th},
\end{equation}
as required by reversibility.  At $\muu=3E/2$ both heats vanish:
\begin{equation}
q_h^{\rm st}=q_c^{\rm st}=0,
\qquad
W_{\rm st}=3\fs=\frac{3E}{2}=\muu_*.
\label{athermalpoint}
\end{equation}
Thus the compensation point is a temperature-neutral reversible chemical stall: the hypothetical reversible cycle converts chemical free energy into an equal amount of mechanical work with no net heat exchange.  The actual steady power remains zero because $J=0$ at stall.

For $\muu<3E/2$, increasing $\Th$ raises the stall force and increasing $\Tc$ lowers it, as in a heat engine.  For $\muu>3E/2$, the signs reverse because chemical driving dominates the thermal contribution.  This inversion is an exact consequence of the mixed affinity.

\section{Closed-form current and an exact chemical-acceleration theorem}
\label{sec:current}

The Metropolis form allows additional exact simplification.  First consider the genuine hybrid heat-engine sector
\begin{equation}
q_h=E+f-\muu\ge0,
\qquad
q_c=E-2f\ge0.
\label{hybridsector}
\end{equation}
Then $\delta_1>0$, $\delta_2\ge0$, and $\delta_3<0$.  Define
\begin{align}
a&=e^{-(E+f)/\Tc},\qquad
b=e^{-(E+f-\muu)/\Th},\notag\\
c&=e^{-(2E-f)/\Tc}.
\label{abc}
\end{align}
The rates reduce to
\begin{equation}
(u_1,u_2,u_3)=\Gc(a,b,1),
\qquad
(w_1,w_2,w_3)=\Gc(1,1,c).
\end{equation}
Substitution into Eq.~(\ref{Jexact}) gives the fully explicit current
\begin{equation}
\boxed{
\frac{J}{\Gc}=\frac{ab-c}{ab+2a+b+bc+2+2c}.}
\label{Jabc}
\end{equation}
The numerator is positive precisely when $ab>c$, equivalently $\mathcal A>0$.

\subsection{Strict monotonicity with chemical affinity}

Only $b$ depends on $\muu$ in Eq.~(\ref{Jabc}), with
\begin{equation}
\frac{\partial b}{\partial\muu}=\frac{b}{\Th}>0.
\end{equation}
Throughout this section and the next two it is convenient to use the
dimensionless denominator
\begin{equation}
\tilde D=ab+2a+b+bc+2+2c=D/\Gc^{2},
\label{Dtilde}
\end{equation}
which is not to be confused with the dimensionful $D$ of Eq.~(\ref{Ddef}).
Direct differentiation gives
\begin{align}
\frac{\partial (J/\Gc)}{\partial b}
&=\frac{a\tilde D-(ab-c)(a+1+c)}{\tilde D^{2}}\notag\\
&=\frac{(2a+c)(1+a+c)}{\tilde D^{2}}>0.
\end{align}
Therefore
\begin{equation}
\boxed{
\frac{\partial J}{\partial\muu}
=\frac{\Gc b}{\Th}
\frac{(2a+c)(1+a+c)}{\tilde D^{2}}>0.}
\label{dJdmuplus}
\end{equation}
This proves that, at fixed $E,f,\Th,\Tc$, chemical free energy strictly accelerates the motor throughout the $q_h\ge0$ Metropolis branch.

The result survives the saturation threshold $q_h=0$.  For $q_h<0$, define
\begin{equation}
d=e^{(E+f-\muu)/\Th}\in(0,1].
\end{equation}
Then $u_2=\Gc$, $w_2=\Gc d$, and
\begin{equation}
\frac{J}{\Gc}=\frac{a-dc}{2a+ad+2+2c+d+dc}.
\label{Jminus}
\end{equation}
Since $\partial d/\partial\muu=-d/\Th$, differentiation yields
\begin{equation}
\boxed{
\frac{\partial J}{\partial\muu}
=\frac{\Gc d}{\Th}
\frac{(a+2c)(1+a+c)}{\tilde D_-^{2}}>0,}
\label{dJdmuminus}
\end{equation}
where $\tilde D_-=2a+ad+2+2c+d+dc$ is the corresponding dimensionless denominator of Eq.~(\ref{Jminus}).  Thus the current is monotonically increasing with fuel affinity on both sides of the Metropolis kink.

\subsection{Finite kinetic ceiling}

\begin{figure}[t]
\includegraphics[width=\columnwidth]{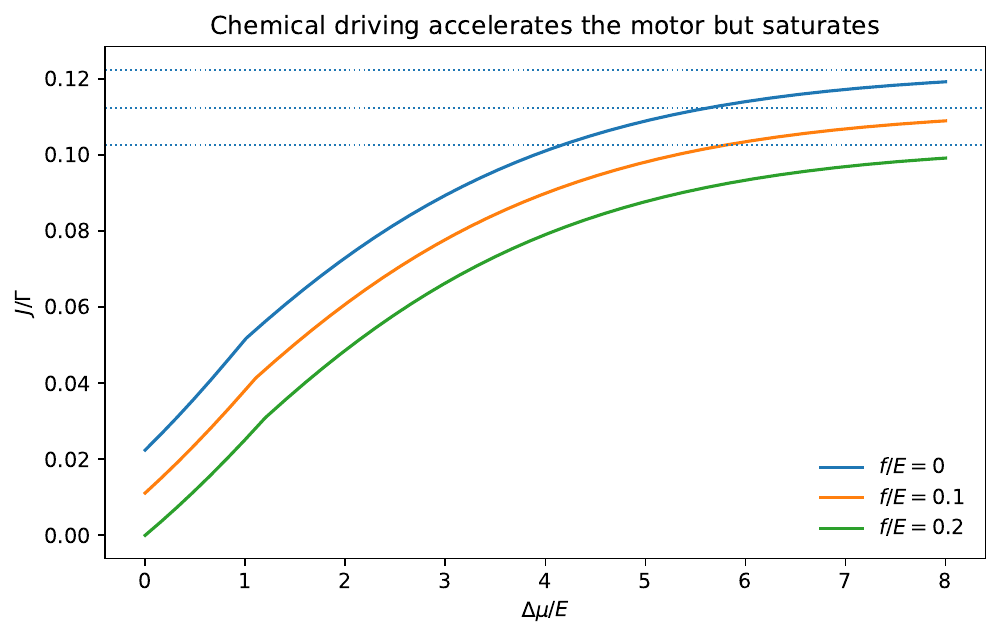}
\caption{Exact stationary current versus chemical affinity for several loads ($E=\Tc=1$, $\Th=2$, $\Gc=1$).  The dotted horizontal lines are the strong-chemical-driving limits in Eq.~(\ref{Jsat}).}
\label{fig:saturation}
\end{figure}

Although $J$ increases monotonically with $\muu$, it does not diverge.  In the strong-driving limit $\muu\to\infty$, $d\to0$ and Eq.~(\ref{Jminus}) gives
\begin{equation}
\boxed{
J_{\infty}(E,f)=\Gc\frac{a}{2(1+a+c)}.}
\label{Jsat}
\end{equation}
Chemical driving makes the hot transition effectively irreversible, but the two cold links remain finite kinetic bottlenecks.  The speed therefore saturates, as shown in Fig.~\ref{fig:saturation}.  At fixed positive load the power $\dot W=3fJ$ inherits the same monotonic chemical enhancement and saturation.

\section{Exact operating modes}
\label{sec:modes}

\begin{figure}[t]
\includegraphics[width=\columnwidth]{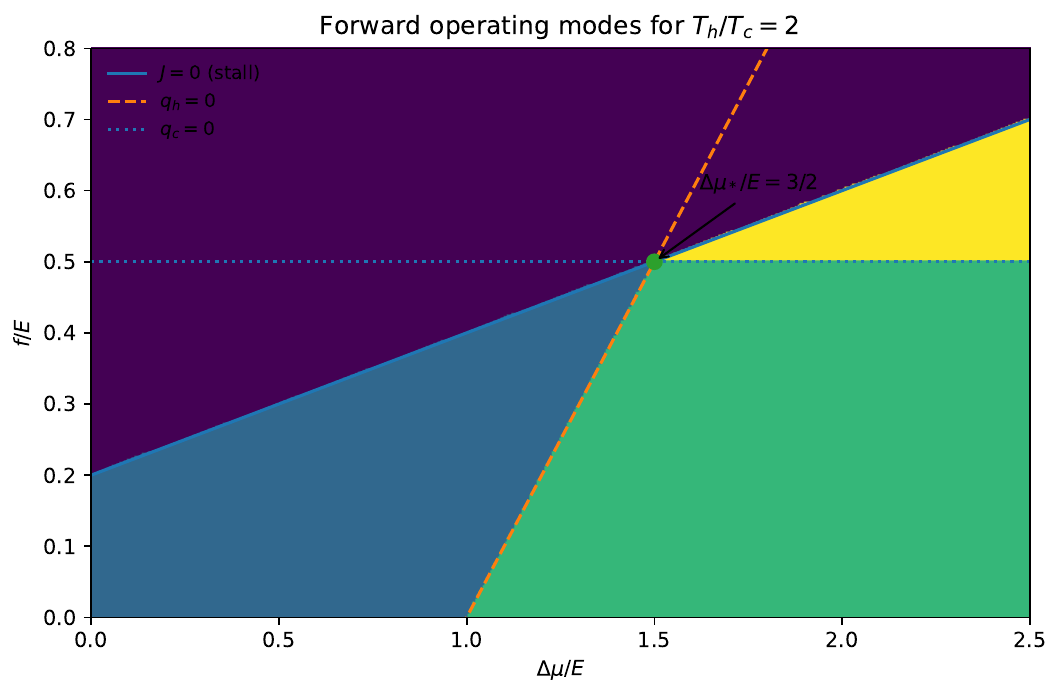}
\caption{Exact operating-mode map for $\Th/\Tc=2$.  The solid line is stall, the dashed line is $q_h=0$, and the dotted horizontal line is $q_c=0$.  The three forward-running colored regions correspond to a hybrid heat engine ($q_h>0,q_c>0$), a chemical motor dissipating into both baths ($q_h<0,q_c>0$), and a chemical motor that also refrigerates the cold bath ($q_h<0,q_c<0$).}
\label{fig:modes}
\end{figure}

For a clockwise cycle three quantities determine the thermodynamic mode:
\begin{equation}
W=3f,\qquad q_h=E+f-\muu,\qquad q_c=E-2f.
\end{equation}
Their boundaries are
\begin{align}
J=0&\Longleftrightarrow f=\fs,\label{boundstall}\\
q_h=0&\Longleftrightarrow f=\muu-E,\label{boundqh}\\
q_c=0&\Longleftrightarrow f=E/2.
\label{boundqc}
\end{align}
The relative positions of these lines are controlled by $\muu/E$.  Two exact identities are
\begin{align}
\fs-\frac{E}{2}
&=\frac{\Tc(2\muu-3E)}{2(2\Th+\Tc)},\label{fsminusqc}\\
\fs-(\muu-E)
&=\frac{\Th(3E-2\muu)}{2\Th+\Tc}.
\label{fsminusqh}
\end{align}
These formulas produce the mode structure in Fig.~\ref{fig:modes}.

For $0\le\muu<E$, every forward-running state $0\le f<\fs$ has $q_h>0$ and $q_c>0$: the machine is a genuine hybrid chemo--thermal heat engine.  For $E<\muu<3E/2$, low loads $0\le f<\muu-E$ give $q_h<0$ and $q_c>0$, so fuel drives both mechanical work and heat dissipation into the hot bath; at larger loads $\muu-E<f<\fs$ the device again draws hot heat and operates as a hybrid engine.  At $\muu=3E/2$ the forward sector ends exactly at $f=E/2$.  For $\muu>3E/2$, $q_h<0$ throughout the forward sector.  When $f<E/2$, chemical free energy produces work while heat is rejected to both reservoirs.  When
\begin{equation}
E/2<f<\fs,
\end{equation}
we have $q_h<0$ and $q_c<0$: the machine absorbs heat from the cold reservoir, rejects heat to the hot reservoir, and simultaneously delivers mechanical work.  The chemical source powers both refrigeration and work extraction.

\section{Power optimization and the speed--load trade-off}
\label{sec:power}

The mechanical power is
\begin{equation}
\dot W(f)=3fJ(f).
\label{powerdef}
\end{equation}
For a forward engine, $\dot W=0$ both at $f=0$ and at $f=\fs$, so at least one interior maximum exists whenever $J>0$ in between.  In the hybrid heat-engine branch, Eqs.~(\ref{Jabc}) and (\ref{abc}) make the stationarity equation completely explicit.

\begin{figure}[t]
\includegraphics[width=\columnwidth]{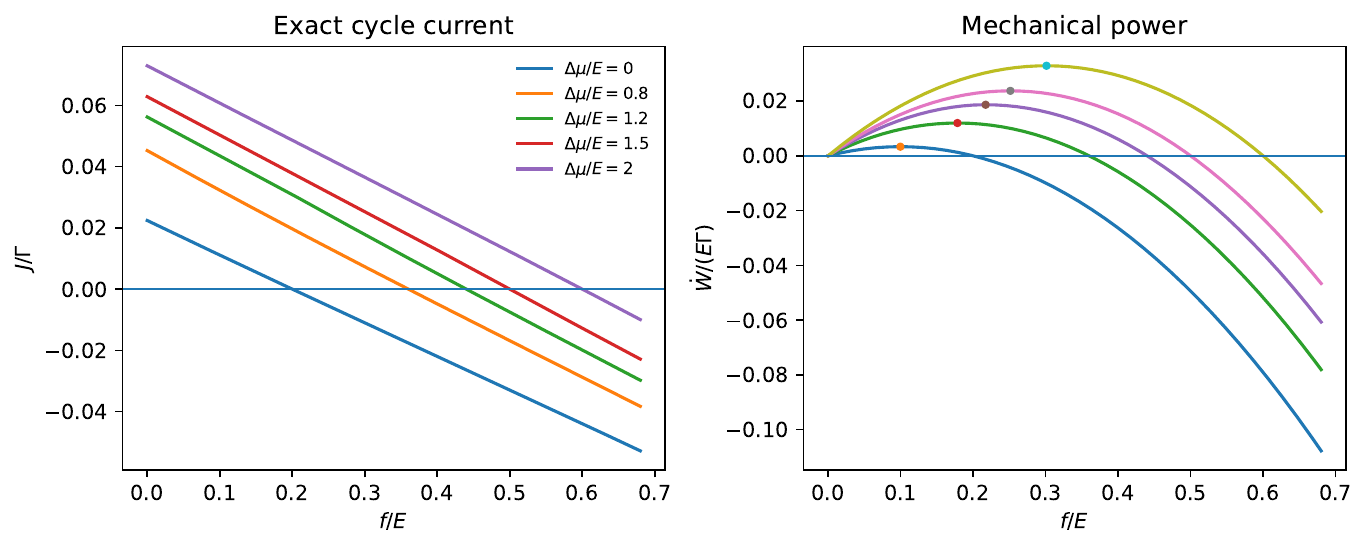}
\caption{Exact current and mechanical power versus load for several chemical affinities ($E=\Tc=1$, $\Th=2$, $\Gc=1$).  Chemical driving moves stall to larger load and increases the maximum power.  The dots indicate numerical solutions of the exact stationarity condition (\ref{fmp_exact}).}
\label{fig:currentpower}
\end{figure}

Write
\begin{equation}
N=ab-c,
\qquad
\tilde D=ab+2a+b+bc+2+2c,
\end{equation}
so that $J=\Gc N/\tilde D$, with $\tilde D$ as in Eq.~(\ref{Dtilde}).  Differentiation with respect to load gives
\begin{align}
N_f&=-(\beta_c+\beta_h)ab-\beta_c c,\label{Nf}\\
\tilde D_f&=-(\beta_c+\beta_h)ab-2\beta_ca-\beta_hb\notag\\
&\quad +(\beta_c-\beta_h)bc+2\beta_cc,
\label{Df}
\end{align}
where $\beta_c=1/\Tc$ and $\beta_h=1/\Th$.  The exact maximum-power load satisfies
\begin{equation}
\frac{d\dot W}{df}=0
\quad\Longleftrightarrow\quad
\boxed{
\frac1f+\frac{N_f}{N}-\frac{\tilde D_f}{\tilde D}=0.}
\label{fmp_exact}
\end{equation}
This is a one-dimensional nonlinear equation with no approximation.  Its solution is readily obtained numerically and is shown in Fig.~\ref{fig:currentpower}.

\subsection{Linear-response limit}

Near equilibrium, the current is linear in the cycle affinity,
\begin{equation}
J=L\mathcal A+O(\mathcal A^2),
\label{linearJ}
\end{equation}
with a positive kinetic coefficient $L$.  To leading order, $L$ may be evaluated at equilibrium and treated as constant while the load changes the affinity.  Using Eq.~(\ref{Astallform}),
\begin{equation}
J\simeq LC_f(\fs-f).
\end{equation}
Hence
\begin{equation}
\dot W\simeq3LC_f f(\fs-f),
\end{equation}
and the maximum occurs at
\begin{equation}
\boxed{f_{\rm mp}^{\rm LR}=\frac{\fs}{2}.}
\label{fmpLR}
\end{equation}
The exact numerical curves in Fig.~\ref{fig:currentpower} lie close to this value for the displayed parameters, but Eq.~(\ref{fmpLR}) should not be interpreted as an exact finite-driving theorem.

A compact result follows for the exergy efficiency.  From the reversible identity at stall,
\begin{equation}
3\fs=\muu+\eta_C(E+\fs-\muu).
\end{equation}
The denominator of Eq.~(\ref{etax}) is linear in $f$.  Evaluating at $f=\fs/2$ gives
\begin{equation}
\boxed{
\eta_{x,\rm mp}^{\rm LR}=\frac{3}{6-\eta_C}.}
\label{etaxmp}
\end{equation}
For the isothermal chemical-motor limit $\eta_C=0$, this reduces to $1/2$.

\section{Optimization with respect to the barrier scale}
\label{sec:barrier}

The energy scale $E$ plays two roles: it creates the thermal rectification affinity and it also suppresses transition rates.  Consequently, current need not increase monotonically with $E$.  In the $q_h>0$ branch, differentiating Eq.~(\ref{Jabc}) at fixed $f,\muu,\Th,\Tc$ gives
\begin{equation}
\frac{\partial J}{\partial E}=\Gc\frac{N_E\tilde D-N\tilde D_E}{\tilde D^{2}},
\end{equation}
with
\begin{align}
N_E&=-(\beta_c+\beta_h)ab+2\beta_cc,\label{NE}\\
\tilde D_E&=-(\beta_c+\beta_h)ab-2\beta_ca-\beta_hb\notag\\
&\quad-(\beta_h+2\beta_c)bc-4\beta_cc.
\label{DE}
\end{align}
Thus an interior speed-optimal barrier obeys the exact equation
\begin{equation}
\boxed{
\frac{N_E}{N}=\frac{\tilde D_E}{\tilde D}.}
\label{Eopt}
\end{equation}
For fixed load, maximizing power with respect to $E$ gives the same condition because $\dot W=3fJ$.

\begin{figure}[t]
\includegraphics[width=\columnwidth]{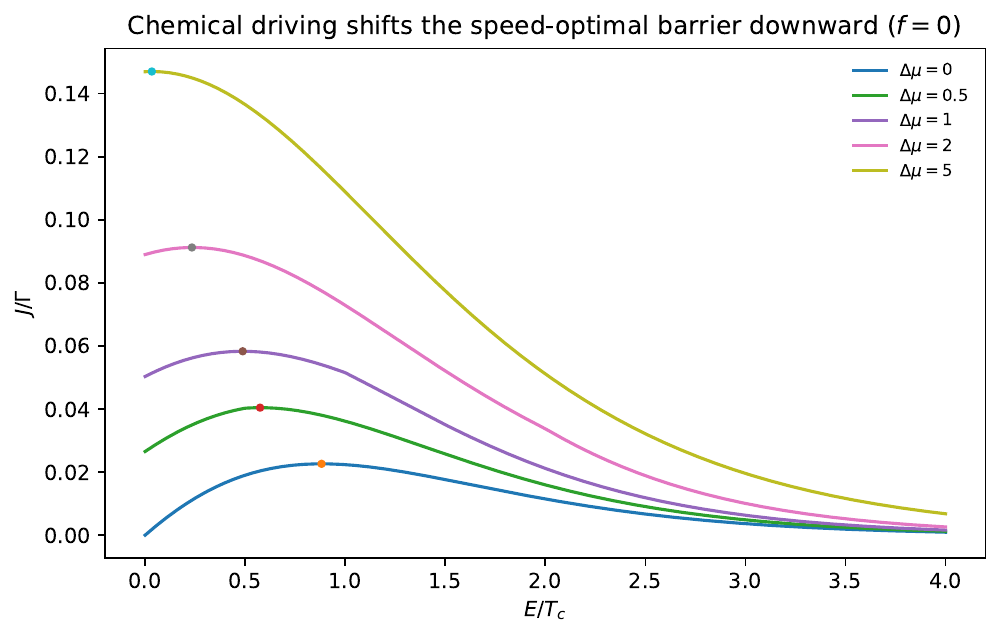}
\caption{Exact current versus barrier scale at zero load for several chemical affinities ($\Tc=1$, $\Th=2$, $\Gc=1$).  Fuel driving shifts the speed-optimal barrier to smaller $E$; under strong chemical driving the thermal barrier ceases to be the primary source of directionality.}
\label{fig:barrier}
\end{figure}

The chemical source qualitatively changes this optimization.  In the passive case $\muu=0$ and zero load, $J\to0$ as $E\to0$ because the thermal rectifier disappears, while $J\to0$ as $E\to\infty$ because activated transitions freeze; therefore an interior maximum is unavoidable.  The same competition between barrier-induced rectification and barrier-induced kinetic suppression governs thermally activated crossing rates more generally \cite{Asfaw2010ThermallyActivatedBarrier,Asfaw2012ExploringDynamicsDimer,Asfaw2014ThermallyActivatedBarrier,Aragie2014ImpurityDiffusionHarmonic,Abebe2022ThermallyActivatedDiffusion}, and it is what an added chemical affinity partially bypasses.  With chemical driving, motion can persist even when the thermal barrier is very small, and the optimal $E$ shifts downward as $\muu$ increases.  The full Metropolis-rate result, including branch changes, is shown in Fig.~\ref{fig:barrier}.

\section{Where the chemical reaction occurs: a general gating theorem}
\label{sec:gating}

The dependence on gating temperature can be derived without specializing to three states.  Consider an arbitrary unicyclic $N$-state motor.  On forward edge $i$, let the internal-energy change be $\Delta\varepsilon_i$, the forward mechanical displacement be $d_i$, the bath temperature be $T_i$, and the stoichiometric number of fuel molecules consumed be $\nu_i$.  Local detailed balance reads
\begin{equation}
\ln\frac{u_i}{w_i}
=-\frac{\Delta\varepsilon_i+fd_i-\nu_i\muu}{T_i}.
\label{generalLDB}
\end{equation}
Summing around the cycle gives
\begin{equation}
\mathcal A
=-\sum_i\frac{\Delta\varepsilon_i}{T_i}
-f\sum_i\frac{d_i}{T_i}
+\muu\sum_i\frac{\nu_i}{T_i}.
\label{generalA}
\end{equation}
Define
\begin{align}
\mathcal A_{\rm th}&=-\sum_i\frac{\Delta\varepsilon_i}{T_i},\notag\\
C_{\rm mech}&=\sum_i\frac{d_i}{T_i},\qquad
C_{\rm chem}=\sum_i\frac{\nu_i}{T_i}.
\end{align}
Then
\begin{equation}
\boxed{
\fs=\frac{\mathcal A_{\rm th}+\muu C_{\rm chem}}{C_{\rm mech}}.}
\label{generalstall}
\end{equation}
For a single fuel event with $\nu=1$ on a link at temperature $T_{\rm ch}$,
\begin{equation}
\boxed{
\frac{\partial\fs}{\partial\muu}=\frac{1}{T_{\rm ch}C_{\rm mech}}.}
\label{chemleverage}
\end{equation}
This is a general chemical-leverage theorem for unicyclic nonisothermal motors: the same chemical free energy produces a larger dimensionless cycle bias when applied to a colder link.

\begin{figure}[t]
\includegraphics[width=\columnwidth]{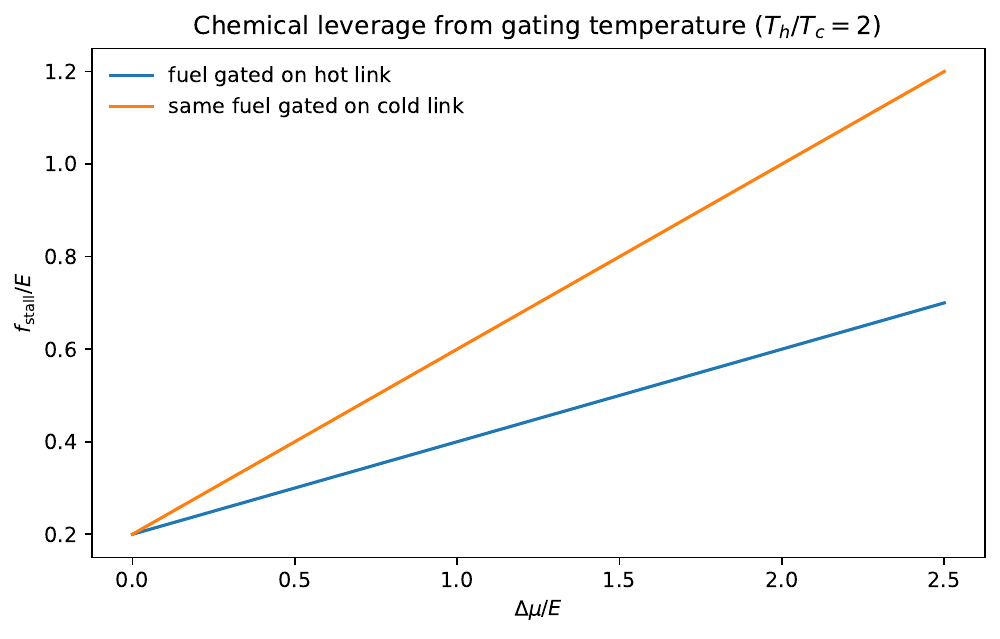}
\caption{Exact stall force for hot-link and cold-link chemical gating.  For the same $\muu$, coupling the chemical event to the colder link gives a larger stall shift by the factor $\Th/\Tc$ in Eq.~(\ref{leverageratio}).}
\label{fig:gating}
\end{figure}

For the present three-state model, $d_i=1$ and
\begin{equation}
C_{\rm mech}=\frac2{\Tc}+\frac1{\Th}.
\end{equation}
Hot-link gating gives Eq.~(\ref{fstall}).  If instead the identical chemical event is coupled to one of the cold links,
\begin{equation}
\boxed{
\fs^{\rm cold}=\frac{E(\Th-\Tc)+\Th\muu}{2\Th+\Tc}.}
\label{fscold}
\end{equation}
The ratio of chemical stall sensitivities is
\begin{equation}
\boxed{
\frac{(\partial\fs/\partial\muu)_{\rm cold}}
{(\partial\fs/\partial\muu)_{\rm hot}}
=\frac{\Th}{\Tc}.}
\label{leverageratio}
\end{equation}
Figure~\ref{fig:gating} illustrates this amplification.

\section{Fluctuation relation for net cycle count}

The affinity also controls rare backward cycles.  Let $n(t)$ denote the net number of completed clockwise windings in a long observation time.  For a stationary unicyclic Markov process, the large-deviation function obeys the Gallavotti--Cohen symmetry associated with the cycle affinity.  Equivalently, at the exponential level in long time,
\begin{equation}
\frac{P_t(n)}{P_t(-n)}\asymp e^{n\mathcal A}.
\label{FTcycle}
\end{equation}
Thus thermal and chemical driving enter current fluctuations only through the same total affinity that controls stall.  At $\mathcal A=0$ the clockwise and counterclockwise winding statistics are symmetric; away from stall their exponential asymmetry is shifted by $\muu/\Th$.  Noise-induced transitions and noise-assisted transport in related stochastic systems display a similar sensitivity to the driving asymmetry \cite{Asfaw2010StochasticResonanceFlexible,Asfaw2011NoiseCreatedBistability,Duki2018StochasticResonanceFirst,Birhanu2021StochasticResonatorLayered,Aragie2022NoiseFormedTriple}.

\section{Minimal multicyclic extension: irreversible mechanical stall}
\label{sec:multicycle}

The exact reversibility of unicyclic stall is mathematically clean but physically restrictive.  Molecular motors can continue consuming fuel while mechanically stalled, and multicyclic kinetic schemes are required to represent such futile turnover \cite{Mugnai2020}.  We therefore add the minimal topological extension needed to separate mechanical stall from thermodynamic reversibility.

Attach a three-state futile chemical loop $2\to4\to5\to2$ to state 2 of the productive mechanical cycle.  Denote its clockwise rates by $a_1,a_2,a_3$, its reverse rates by $b_1,b_2,b_3$, and its affinity by
\begin{equation}
\mathcal A_f=\ln\frac{a_1a_2a_3}{b_1b_2b_3}.
\end{equation}
The productive cycle retains affinity $\mathcal A_m=\mathcal A$ from Eq.~(\ref{affinity}).  The two cycles share only the articulation state 2 and no edge, so every edge of
the network belongs to exactly one cycle.  Consequently the net current across
any edge of the productive ring equals the mechanical cycle flux $J_m$, and the
net current across any edge of the futile ring equals the futile cycle flux
$J_f$; the two cycle currents may therefore be treated independently.

\begin{figure}[t]
\includegraphics[width=\columnwidth]{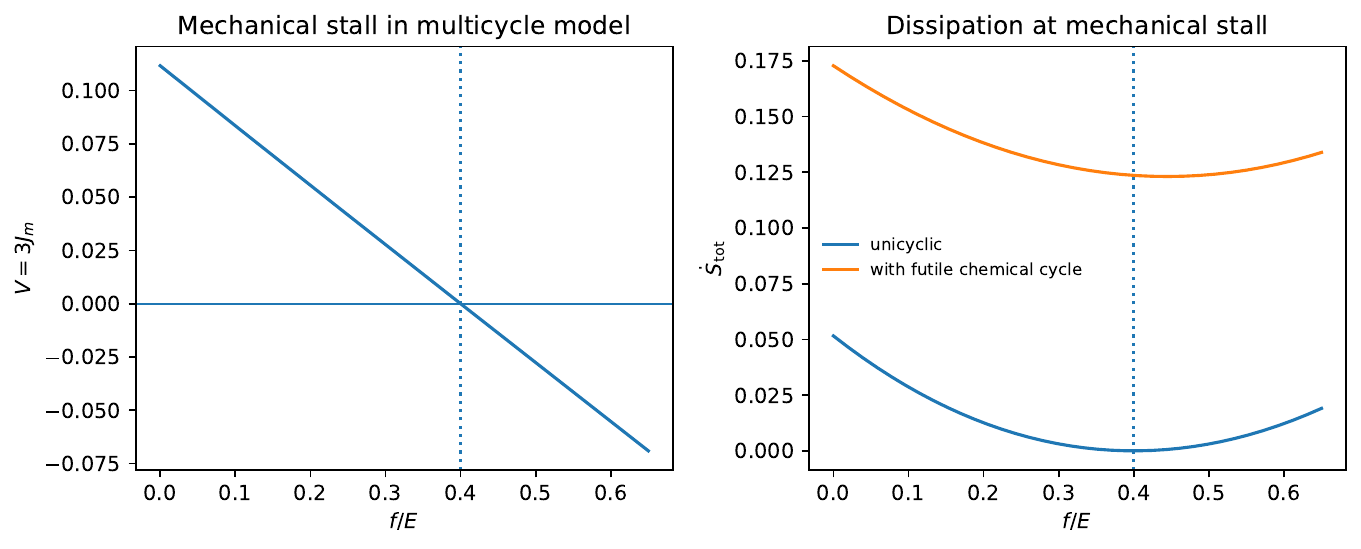}
\caption{Minimal multicycle extension for $E=\Tc=1$, $\Th=2$, $\muu=1$, with a futile-cycle affinity $\mathcal A_f=2$.  The mechanical velocity vanishes at the same productive-cycle stall load, but total entropy production remains positive because the futile chemical cycle continues to turn.}
\label{fig:multicycle}
\end{figure}

The vanishing of the mechanical current at stall now follows from cycle-flux
theory rather than from an assumption of detailed balance.  For a Markov network
decomposed into cycles, the net flux of a cycle $\kappa$ is proportional to the
difference of its forward and reverse rate products \cite{Schnakenberg1976},
\begin{equation}
J_\kappa\propto\prod_{i\in\kappa}u_i-\prod_{i\in\kappa}w_i
=\left(\prod_{i\in\kappa}w_i\right)\left(e^{\mathcal A_\kappa}-1\right),
\label{cycleflux}
\end{equation}
with a strictly positive proportionality factor built from spanning-tree
products of the remaining network.  It is essential that this statement applies
cycle by cycle: a vanishing cycle affinity forces the corresponding cycle flux
to vanish even when other cycles of the same network remain driven.  At the
mechanical stall load $f=\fs$ we have $\mathcal A_m=0$, so Eq.~(\ref{cycleflux})
gives
\begin{equation}
J_m=0,
\qquad
V=3J_m=0.
\label{Jmstall}
\end{equation}
Because the productive edges carry no other cycle, each of them separately
satisfies detailed balance at this load; this is a consequence of
Eq.~(\ref{Jmstall}), not an assumption used to derive it.  Note also that
$\mathcal A_m$ is unchanged by the attachment of the futile loop, since the loop
adds no edge to the productive ring, so the stall load remains exactly
Eq.~(\ref{fstall}).
The futile current, however, need not vanish.  Solving the three steady current equations on the loop for a given shared-state probability $p_2>0$ gives
\begin{equation}
\boxed{
J_f=p_2\frac{a_1a_2a_3-b_1b_2b_3}
{a_2a_3+a_3b_1+b_1b_2}.}
\label{Jfexact}
\end{equation}
Therefore $J_f\ne0$ whenever $\mathcal A_f\ne0$.  Schnakenberg cycle decomposition gives
\begin{equation}
\ep=J_m\mathcal A_m+J_f\mathcal A_f.
\label{epmulti}
\end{equation}
At mechanical stall,
\begin{equation}
\boxed{
V=0,
\qquad
\ep=J_f\mathcal A_f>0.}
\label{irrevstall}
\end{equation}
Thus one additional independent cycle is sufficient to obtain an idling, fuel-consuming, dissipative motor.  Figure~\ref{fig:multicycle} compares the vanishing unicyclic entropy production at stall with the finite multicyclic value.

\section{Relation to purely thermal Brownian engines and molecular motors}
\label{sec:relation}

The detailed quasistatic limits were derived in Sec.~\ref{sec:quasistatic}.  They show explicitly that the same stochastic engine interpolates continuously between a passive Brownian heat engine and a tight-coupling chemical motor.  For $\muu=0$, the reversible heat-conversion efficiency approaches $\eta_C$; for $\Th=\Tc$, reversible stall satisfies $3\fs=\muu$ and the chemical conversion efficiency approaches unity.  With both resources present, the common reversible statement is instead the exergy relation $W_{\rm rev}=\muu+\eta_Cq_h$.

The present framework therefore joins Brownian heat-engine thermodynamics \cite{Asfaw2005EnergeticsSimpleMicroscopic,Asfaw2008ModelingEfficientBrownian,Taye2015ExactAnalyticalThermodynamic,Taye2022ExactTimeDependent2,Taye2024ExactTimeDependent,Taye2025CurzonAhlbornType} to chemical molecular-motor thermodynamics \cite{Magnasco1994,Julicher1997,Seifert2011,SchmiedlSeifert2007,RaoEsposito2016,Mugnai2020}.  Unlike a generic molecular-motor model, however, the two thermal reservoirs remain explicit.  This makes it possible to identify the temperature-neutral compensation point, chemical gating leverage, and combined motor--refrigerator regime analytically.

\section{Discussion}

Several distinctions are important for interpreting the results.

First, the exact stall force depends only on local-detailed-balance ratios, whereas finite current and power also depend on the kinetic convention used to realize those ratios.  Equation~(\ref{fstall}) is therefore more robust than the detailed current formula (\ref{Jabc}).  Replacing Metropolis rates by another locally detailed-balanced rate family leaves the affinity and stall unchanged but generally changes the magnitude of $J$, the maximum-power load, and the barrier optimum.

Second, chemical free energy and heat are not thermodynamically interchangeable.  The generalized reversible identity $W=\muu+\eta_Cq_h$ shows that chemical free energy carries unit exergy, while heat carries only the Carnot fraction.  Consequently, efficiencies defined using heat alone or fuel alone can exceed familiar single-resource bounds without violating the second law.  The exergy efficiency in Eq.~(\ref{etax}) is the appropriate common measure.

Third, the compensation point $\muu_*=3E/2$ is not an equilibrium point of the reservoirs; $\Th$ and $\Tc$ may remain different.  Rather, it is a point where the thermal energy transfers of the reversible cycle vanish individually.  It is therefore a cancellation in cycle energetics, not a removal of the imposed temperature difference.

Fourth, the monotonic chemical-acceleration theorem is kinetic as well as thermodynamic.  The chemical term increases the affinity, but Eq.~(\ref{dJdmuplus}) and Eq.~(\ref{dJdmuminus}) prove more: with Metropolis kinetics it increases the actual current on both sides of the hot-link saturation threshold.  The finite ceiling in Eq.~(\ref{Jsat}) shows why arbitrarily large chemical free energy cannot produce arbitrarily large speed.

Finally, the multicyclic extension clarifies the meaning of stall.  In a unicyclic network, $J=0$ forces detailed balance and zero entropy production.  In a multicyclic motor, only the mechanically productive cycle need stall.  Hidden or futile cycles can continue to consume chemical free energy and dissipate heat, a feature expected in realistic biomolecular machines \cite{Mugnai2020}.  The same figure-eight topology, in which one ring carries a mechanical current while an attached ring turns over chemically, is the natural starting point for finite-time and first-passage questions---how long the productive ring takes to complete a winding, and how the futile ring reshapes that distribution---which have been studied for related stochastic systems \cite{Duki2016FirstPassageTime,Duki2018StochasticResonanceFirst2,Taye2026StochasticFirstPassage} and for time-dependent driving of Brownian engines \cite{Taye2023TimeDependentThermodynamic,Taye2025ThermodynamicRelationsTime,Taye2025EntropyProductionThermodynamic,Taye2026UniversalThermodynamicInequality}.  We leave such extensions of the present exactly solvable model to future work.

\section{Conclusion}

We have constructed a detailed exact chemo--thermal extension of the three-state Metropolis Brownian heat engine.  The full stationary probabilities, current, cycle affinity, stall load, heat flows, entropy production, exergy balance, operating modes, and optimization conditions follow analytically from local detailed balance.

The central exact results are
\begin{equation}
\mathcal A=E(\Tc^{-1}-\Th^{-1})+\muu/\Th-f(2/\Tc+1/\Th),
\end{equation}
\begin{equation}
\fs=\frac{E(\Th-\Tc)+\Tc\muu}{2\Th+\Tc},
\end{equation}
and
\begin{equation}
\muu+\eta_Cq_h-W=\Tc\Delta S_{\rm cyc}.
\end{equation}
The combination of thermal and chemical driving produces the temperature-neutral compensation point $\muu_*=3E/2$, where $\fs=E/2$ and the reversible cycle exchanges no net heat with either bath.  Chemical driving monotonically increases the Metropolis current but only to a finite kinetic ceiling.  Maximum power is determined by an exact one-dimensional stationarity equation and approaches half-stall loading in linear response.  The location of chemical gating supplies an additional control variable, with a universal leverage proportional to the inverse local temperature.  Finally, a minimal multicyclic extension permits finite fuel consumption and entropy production at zero mechanical velocity.

These results identify a compact exactly solvable platform in which heat-engine physics, chemical-motor physics, refrigeration, finite-power optimization, and hidden dissipation can be studied within one stochastic network.

\begin{acknowledgments}
The author thanks the Science Division of West Los Angeles College for support.
\end{acknowledgments}

\section*{Data availability}
All results in this paper are analytical.  The scripts used to evaluate the exact expressions and to generate the figures are available from the author upon reasonable request.

\appendix

\section{Derivation of the stationary probabilities}
\label{app:stationary}

At stationarity Eqs.~(\ref{me1})--(\ref{me3}) can be written $\bm M\bm p=0$, with
\begin{equation}
\bm M=
\begin{pmatrix}
-(u_1+w_3)&w_1&u_3\\
u_1&-(w_1+u_2)&w_2\\
w_3&u_2&-(u_3+w_2)
\end{pmatrix}.
\end{equation}
For a continuous-time Markov chain, the stationary weight of a state equals the sum of directed spanning-tree products rooted at that state.  The three rooted weights are
\begin{align}
\tau_1&=u_2u_3+u_3w_1+w_1w_2,\\
\tau_2&=u_1u_3+u_1w_2+w_2w_3,\\
\tau_3&=u_1u_2+u_2w_3+w_1w_3.
\end{align}
Their sum is exactly $D$ in Eq.~(\ref{Ddef}), hence $p_i^s=\tau_i/D$.  Substituting into $J=u_1p_1-w_1p_2$ gives
\begin{align}
JD&=u_1(u_2u_3+u_3w_1+w_1w_2)\notag\\
&\quad-w_1(u_1u_3+u_1w_2+w_2w_3)\notag\\
&=u_1u_2u_3-w_1w_2w_3,
\end{align}
which proves Eq.~(\ref{Jexact}).

\section{Continuity of the current at the Metropolis threshold}

At $q_h=0$, or $\muu=E+f$, the hot forward and backward rates both equal $\Gc$.  In the $q_h\ge0$ representation, $b=1$ and
\begin{equation}
\frac{J}{\Gc}=\frac{a-c}{3a+3c+3}=
\frac{a-c}{3(1+a+c)}.
\end{equation}
In the $q_h<0$ representation, $d=1$ and the same expression follows.  Hence $J$ is continuous through the Metropolis kink, although it need not be differentiable there, because the rate law itself has a cusp.

\section{Exact derivatives used in power optimization}

In the $q_h\ge0$ sector,
\begin{equation}
a_f=-\beta_ca,\qquad b_f=-\beta_hb,\qquad c_f=+\beta_cc.
\end{equation}
Therefore
\begin{align}
N_f&=a_fb+ab_f-c_f
=-(\beta_c+\beta_h)ab-\beta_cc,\\
\tilde D_f&=(a_fb+ab_f)+2a_f+b_f+(b_fc+bc_f)+2c_f,
\end{align}
which reduces to Eq.~(\ref{Df}).  Since
\begin{equation}
\frac{d\dot W}{df}=3\Gc\left[\frac{N}{\tilde D}+f\frac{N_f\tilde D-N\tilde D_f}{\tilde D^{2}}\right],
\end{equation}
setting this derivative to zero and dividing by $fN/\tilde D$ gives Eq.~(\ref{fmp_exact}).

\section{Additional identities at stall}

Starting from Eq.~(\ref{fstall}),
\begin{align}
E+\fs-\muu
&=E-\muu+
\frac{E(\Th-\Tc)+\Tc\muu}{2\Th+\Tc}\notag\\
&=\frac{\Th(3E-2\muu)}{2\Th+\Tc},
\end{align}
which proves Eq.~(\ref{qhstall}).  Similarly,
\begin{align}
E-2\fs
&=\frac{E(2\Th+\Tc)-2E(\Th-\Tc)-2\Tc\muu}{2\Th+\Tc}\notag\\
&=\frac{\Tc(3E-2\muu)}{2\Th+\Tc}.
\end{align}
Their ratio immediately gives $q_c^{\rm st}/\Tc=q_h^{\rm st}/\Th$.

\section{Futile-cycle current in the figure-eight network}

Let $p_2,p_4,p_5$ denote the probabilities on the futile loop and $J_f$ its clockwise current.  The steady current equations are
\begin{align}
J_f&=p_2a_1-p_4b_1,\\
J_f&=p_4a_2-p_5b_2,\\
J_f&=p_5a_3-p_2b_3.
\end{align}
The first equation gives
\begin{equation}
p_4=\frac{p_2a_1-J_f}{b_1},
\end{equation}
and the second gives
\begin{equation}
p_5=\frac{a_2(p_2a_1-J_f)-b_1J_f}{b_1b_2}.
\end{equation}
Substituting this into the third equation and collecting $J_f$ yields
\begin{equation}
J_f(a_2a_3+a_3b_1+b_1b_2)
=p_2(a_1a_2a_3-b_1b_2b_3),
\end{equation}
which proves Eq.~(\ref{Jfexact}).

\section{Numerical evaluation}

All figures use the exact Metropolis definitions (\ref{metu})--(\ref{metw}) rather than branch-specific approximations.  Unless stated otherwise, the dimensionless parameters are $E=1$, $\Tc=1$, $\Th=2$, and $\Gc=1$.  Stationary probabilities in the five-state multicycle model are obtained by replacing one row of the generator by the normalization condition and solving the resulting linear system.  Numerical maximization of power is one-dimensional over the forward interval $0<f<\fs$ and is used only to illustrate the exact analytic stationarity equation (\ref{fmp_exact}).

\end{document}